\documentclass[prl,aps,preprint,superscriptaddress,showpacs]{revtex4}
\usepackage{txfonts}
\usepackage{graphicx}
\begin{document}

\title{ Origin of the inner ring in photoluminescence patterns of
quantum well excitons }

\author{A.\,L. Ivanov}
\affiliation{Department of Physics and Astronomy, Cardiff
University, Cardiff CF24 3YB, United Kingdom}

\author{L.\,E. Smallwood}
\affiliation{Department of Physics and Astronomy, Cardiff
University, Cardiff CF24 3YB, United Kingdom}

\author{A.\,T. Hammack}
\author{Sen Yang}
\author{L.\,V. Butov}
 \affiliation{Department of Physics,
University of California at San Diego, La Jolla, CA 92093-0319}

\author{A.\,C. Gossard}
\affiliation{Materials Department, University of California at Santa
Barbara, Santa Barbara, California 93106-5050}

\date{\today}

\begin{abstract}

In order to explain and model the {\it inner ring} in
photoluminescence (PL) patterns of indirect excitons in GaAs/AlGaAs
quantum wells (QWs), we develop a microscopic approach formulated in
terms of coupled nonlinear equations for the diffusion,
thermalization and optical decay of the particles. The origin of the
inner ring is unambiguously identified: it is due to cooling of
indirect excitons in their propagation from the excitation spot. We
infer that in our high-quality structures the in-plane diffusion
coefficient $D^{\rm (2d)}_{\rm x} \simeq 10-30$\,cm$^2$/s and the
amplitude of the disorder potential $U_0/2 \simeq 0.45$\,meV.

\end{abstract}


\maketitle

More than two decades ago, long-distance diffusion and drift
transport of charge-neutral excitons with a long lifetime was
optically visualized in bulk Si and Cu$_2$O
\cite{Tamor80,Trauernicht84}. In these earlier works, strain
gradient potential traps were used to induce the drift motion. Due
to the low particle concentrations, the transport was described in
terms of a classical picture, with no quantum-statistical corrections,
and the diffusion coefficient was attributed to exciton-phonon
scattering. In-plane propagation of long-lived indirect excitons in
coupled QWs over large distances has also been reported
\cite{Hagn95,Negoita99,Larionov00,Butov02b,Voros05}: in this case
one has a unique possibility to optically map the
quasi-two-dimensional motion of composite bosons. Furthermore, the
density $n_{\rm 2d}$ of indirect excitons can be large enough to
ensure nonclassical population of the ground-energy state, $N_{E=0}
= e^{T_0/T} - 1 \gtrsim 1$, where $T_0 = (\pi \hbar^2 n_{\rm 2d})/
(2 k_{\rm B} M_{\rm x})$ and $T$ are the degeneracy temperature and
exciton temperature, respectively, and $M_{\rm x}$ is the exciton
in-plane translational mass (for a recent review of cold exciton
gases in coupled QWs see Ref.\,\cite{Butov04b}).

One of the most striking features of photoluminescence associated
with indirect excitons is the appearance of two {\it PL rings}
\cite{Butov02b}. While the second, external ring has already been
explained in terms of in-plane spatially separated electrons and
holes \cite{Butov04a,Rapaport04}, the origin of the first, inner ring,
which arises purely due to the transport of indirect excitons,
remained unclear. The underlying physical picture we propose to
explain this ring is that in the optically-pumped area the exciton
temperature $T$ is much larger than the lattice temperature $T_{\rm b}$.
As a result, the optical decay of excitons is suppressed, but while
they diffuse out they cool down and eventually become optically-active,
giving rise to a local increase of the PL signal. Thus the first ring
is a generic feature of cw PL from excitons, direct or indirect, that
are nonresonantly excited in high-quality QWs.

In this Letter we develop a microscopic theory for the long-range
transport, thermalization and optical decay of QW excitons, model
the inner PL ring, show an effective screening of QW disorder for
$n_{\rm 2d} \gtrsim 10^{10}$\,cm$^{-2}$, and analyze the
quantum-statistical corrections. By numerically fitting the PL spectra
we clarify the main scattering channels which contribute to the
diffusion of indirect excitons and evaluate the diffusion coefficient
and amplitude of the QW disorder potential. In the proposed model, the
exciton temperature $T$ is affected by heating due to the optical
excitation, heating due to the LA-phonon assisted conversion of the
mean-field energy into the internal energy, and recombination
heating or cooling due to the optical decay of low-energy excitons.
The mean-field energy of indirect excitons also gives rise to a
potential energy gradient and, therefore, to the in-plane drift motion.

In the experiments, the PL pattern is imaged by a nitrogen-cooled
CCD camera with the spectral selection by an interference filter
adjusted to the indirect exciton energy. Fine adjustment of the
filtering energy is done by rotating two interference filters relative
to the optical axis. As a result, we are able to remove the
low-energy bulk emission that otherwise dominates the spectrum near
the excitation spot \cite{Butov02b}, and observe a two-dimensional
spatial image of the inner ring (see Figs.\,1a, 1c, and 1d).
Note that the inner PL ring can be missed if the bulk emission
is not removed from the PL signal. In Figs.\,1e and 1f we plot the
measured exciton PL in the {\it energy-coordinate} plane. The exciton
energy $E_{\rm PL}$ decreases with increasing distance from the
excitation spot, as detailed below. This results in an arrow-shaped
profile of the exciton PL images in the $E$-$x$ coordinates. The
external ring is also seen at high excitations, both in $x$-$y$
(Fig.\,1a) and $E$-$x$ (Fig.\,1e) coordinates. The excitation
is done by a HeNe laser at 633\,nm (the laser spot is a Gaussian
with FWHM $= 2 \sigma \simeq 6$\,$\mu$m, the excitation power
$P_{\rm ex} = 1-400\,\mu$W). The coupled QW structure with two
$8$\,nm GaAs QWs separated by a $4$\,nm Al$_{0.33}$Ga$_{0.67}$As
barrier is grown by molecular beam epitaxy.

Our approach to the transport, relaxation and PL dynamics of indirect
excitons is formulated in terms of three coupled nonlinear equations:
a quantum diffusion equation for $n_{\rm 2d}$, a thermalization equation
for $T$, and an equation for the optical lifetime $\tau_{\rm opt} = 1/
\Gamma_{\rm opt}$. The quantum-statistical corrections, which enhance
the nonlinear effects, are included in the description.

The nonlinear quantum diffusion equation is given by
\begin{eqnarray}
\label{diff}
\frac{\partial n_{\rm 2d}}{\partial t} = \nabla \Big[
D_{\rm x}^{\rm (2d)} \nabla n_{\rm 2d} &+& \mu_{\rm x}^{\rm (2d)}
n_{\rm 2d} \nabla \left( u_0 n_{\rm 2d} + U_{\rm QW} \right) \Big]
- \Gamma_{\rm opt}n_{\rm 2d}  + \Lambda \, ,
\end{eqnarray}
where $\Gamma_{\rm opt}$, $D_{\rm x}^{\rm (2d)}$, $\mu_{\rm x}^{\rm
(2d)}$, and $\Lambda$ are the radiative decay rate, diffusion
coefficient, mobility, and generation rate of QW excitons,
respectively, $U_{\rm QW} = U_{\rm rand}({\bf r}_{\|})$ is a random
potential due to the QW thickness and alloy fluctuations, and the
operator $\nabla$ is defined in terms of the in-plane coordinate
vector ${\bf r}_{\|}$. The {\it generalized Einstein relationship}
\cite{Ivanov02}, $\mu_{\rm x}^{\rm (2d)} = D_{\rm x}^{\rm (2d)}
[(e^{T_0/T} - 1)/(k_{\rm B}T_0)]$, which yields the classical limit
$\mu_{\rm x}^{\rm (2d)} = D_{\rm x}^{\rm (2d)}/(k_{\rm B}T)$ for $T
\gg T_0$, strongly enhances the nonlinearity of the diffusion
Eq.\,(\ref{diff}) for $T \lesssim T_0$. The positive mean-field
energy $u_0 n_{\rm 2d}$ on the right-hand side (r.h.s.) of
Eq.\,(\ref{diff}) is due to the well-defined dipole-dipole repulsive
interaction of indirect excitons \cite{Yoshioka90,Butov94,Zhu95}.
Here $u_0 = 4 \pi (\mbox{e}^2/\varepsilon_{\rm b}) d_{\rm z}$,
$\varepsilon_{\rm b}$ is the background dielectric constant, and
$d_{\rm z}$ is the separation between electron and hole layers. The
mean-field energy gives rise to the in-plane drift motion with the
velocity ${\bf v}_{\rm drift} = - \mu^{\rm (2d)}_{\rm x} u_0 \nabla
n_{\rm 2d}$. As a result, an effective screening of the disorder
potential $U_{\rm rand}$ by dipole-dipole interacting indirect
excitons builds up with increasing $n_{\rm 2d}$
\cite{Ivanov02,Zimmermann05}: the excitons tend to accumulate near
the minima of $U_{\rm rand}({\bf r}_{\|})$ [local increase of $u_0
n_{\rm 2d}({\bf r}_{\|})$] and avoid the maxima of
$U_{\rm rand}({\bf r}_{\|})$ [local decrease of
$u_0 n_{\rm 2d}({\bf r}_{\|})$]. As we show below, in our high-quality
structures $U^{(0)} = 2 \langle |U_{\rm rand}({\bf r}_{\|})| \rangle
\simeq 0.9$\,meV and the mean-free energy $u_0 n_{\rm 2d}^{(0)} \simeq
1.6$\,meV for $n^{(0)}_{\rm 2d} = 10^{10}$\,cm$^{-2}$, so that at low
exciton temperatures $T \sim 1$\,K the QW disorder is strongly screened
and practically removed for $n_{\rm 2d} \gtrsim n^{(0)}_{\rm 2d}$.

The temperature dynamics of excitons is given by
\begin{eqnarray}
\label{therm}
&&\frac{\partial}{\partial t} T = \left( \frac{\partial T}{\partial t}
\right)_{n_{\rm 2d}} + \ \frac{ S_{\rm pump} + S_{\rm opt} +
S_{\rm d} }{ 2 k_{\rm B} T I_1 - k_{\rm B} T_0 I_2 } \,, \\
\label{thermLA}
&&\left( {\partial T \over \partial t} \right)_{n_{\rm 2d}} \!\!\!\!
= - { 2 \pi \over \tau_{\rm sc} } \left( {T^2 \over T_0} \right)
\big(1 - e^{-T_0/T} \big) \!\! \int_1^{\infty} \!\!\! d \varepsilon \
\varepsilon \sqrt{\varepsilon \over \varepsilon - 1}
\nonumber \\
&& \ \ \ \ \ \ \ \ \ \ \ \ \ \ \ \ \ \ \ \ \ \ \ \
\times \frac{|F_z (a \sqrt{\varepsilon(\varepsilon - 1})
|^2}{(e^{\varepsilon E_0 / k_{\rm B} T_{\rm b}} - 1) } \
{e^{\varepsilon E_0 / k_{\rm B} T_{\rm b}} - e^{\varepsilon E_0 /
k_{\rm B} T} \over (e^{\varepsilon E_0 / k_{\rm B} T} + e^{-T_0/T} -
1) } \, ,
\end{eqnarray}
where $I_1 = (1 - e^{-T_0/T}) \int_0^{\infty} dz [z/(e^z +
e^{-T_0/T} - 1)]$, $I_2 = e^{-T_0/T} \int_0^{\infty} dz [(z
e^z)/(e^z + e^{-T_0/T} - 1)^2]$. Equation (\ref{thermLA}) describes
the LA-phonon assisted thermalization of indirect excitons:
$\tau_{\rm sc} = (\pi^2 \hbar^4 \rho)/(D_{\rm dp}^2 M_{\rm x}^3
v_{\rm s})$ is the characteristic scattering time, $E_0 = 2M_{\rm x}
v_{\rm s}^2$, $v_{\rm s}$ is the longitudinal sound velocity, $\rho$
is the crystal density, and $D_{\rm dp}$ is the deformation potential
of exciton -- LA-phonon interaction. The form-factor $F_z(\chi) =
[\sin(\chi)/\chi] [e^{i \chi}/(1 - \chi^2/\pi^2)]$ refers to a QW
confinement potential and characterizes a spectral band of bulk
LA-phonons which interact with indirect excitons, $a = (d_{\rm QW}
M_{\rm x} v_{\rm s})/\hbar$, and $d_{\rm QW}$ is the QW thickness.
For $1.5\,\mbox{K} \lesssim T \lesssim 6.5\,{\mbox K}$ and
$n_{\rm 2d} = 10^{10}$\,cm$^{-2}$, relevant to our experiments,
Eq.\,(\ref{thermLA}) yields a thermalization time $0.15\,\mbox{ns}
\gtrsim \tau_{\rm th} \gtrsim 0.03$\,ns. The strong increase of
$\tau_{\rm th}$ with decreasing $T$ is due to classical slowing down
of the relaxation kinetics, which occurs at $k_{\rm B} T \sim E_0$
(excitons with energies $E \leq E_0/4$ cannot emit LA-phonons)
\cite{Ivanov99}.

\begin{figure}
\includegraphics*[width=7.7cm]{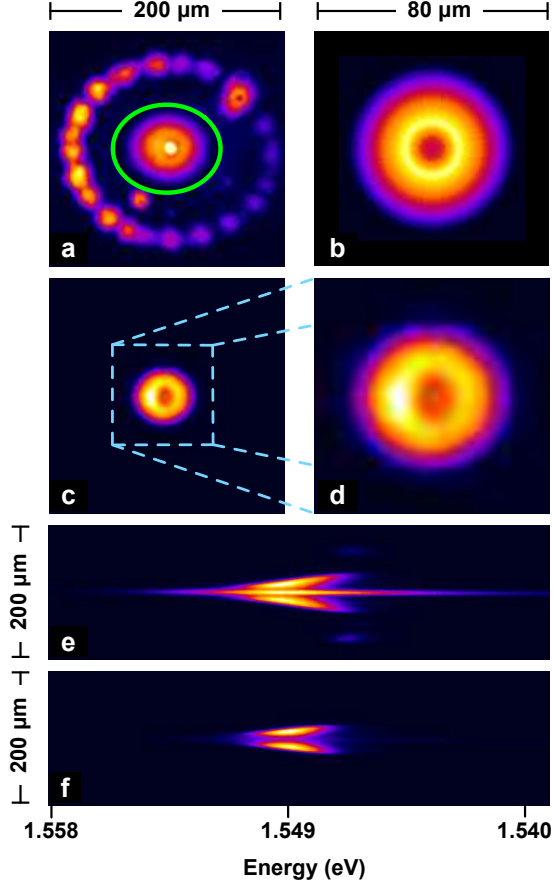}
\caption{(a), (c)-(f) Experimental and (b) calculated patterns of
the PL signal from indirect excitons. In (a), the PL intensity in
the area within the green circle is reduced by a constant factor
for better visualization. A bright spot in the middle of the inner
ring in (a) is due to residual bulk emission. (e) and (f) Image of
the PL signal in the $E$-$x$ coordinates. The external PL ring
is also seen for high excitations, both in (a) $x$-$y$ and (e)
$E$-$x$ coordinates. For (a) and (e) the excitation power is
$P_{\rm ex}= 250\,\mu$W and for (b)-(d) and (f) $P_{\rm ex} =
75\,\mu$W, respectively, and $T_{\rm b} = 1.5\,$K. }
\label{Fig1}
\end{figure}

Heating of indirect excitons by the laser pulse is given by the term
$S_{\rm pump} = (E_{\rm inc} - k_{\rm B} T I_2) \Lambda_{T_0} > 0$
on the r.h.s. of Eq.\,(\ref{therm}). Here, $\Lambda_{T_0} = [(\pi
\hbar^2)/ (2 k_{\rm B} M_{\rm x})] \Lambda(t,{\bf r}_{\|})$. The
generation of indirect excitons is a secondary process, mainly due to
quantum tunnelling of photoexcited direct excitons to the energetically
more favorable states of indirect excitons. The excess energy
$E_{\rm inc}$ of a created (incoming) indirect exciton is large: it
exceeds the energy splitting between the direct and indirect excitons,
which is about 20\,meV. For the highest generation rates used in the
experiments, the exciton temperature $T^{\rm max} \simeq 6.4$\,K at
the laser spot centre is much larger than $T_{\rm b} \simeq 1.5$\,K.

Recombination heating or cooling of QW excitons is given by the term
$S_{\rm opt} = [k_{\rm B}T I_2 \Gamma_{\rm opt} - E_{\gamma}
\Gamma_{\rm opt}^{\rm E}] T_0$ on the r.h.s. of Eq.\,(\ref{therm}),
where
\begin{equation}
\label{opt}
\Gamma_{\rm opt} = { 1 \over 2 \tau_{\rm R}}
\Bigg( {E_{\gamma} \over k_{\rm B} T_0 } \Bigg) \int_0^{1} \!\! { 1
+ z^2 \over B e^{- z^2 E_{\gamma}/k_{\rm B} T} - 1 }\,\,dz
\end{equation}
and $\Gamma_{\rm opt}^{\rm E} = [E_{\gamma}/(2 \tau_{\rm R} k_{\rm B}
T_0)] \int_0^{1}  [(1 - z^4) / (Be^{- z^2 E_{\gamma}/k_{\rm B} T} -
1)]\,\,dz$ are the optical decay rates for the concentration and energy
density of indirect excitons, respectively, $\tau_{\rm R}$ is the
intrinsic radiative lifetime of indirect excitons, $E_{\gamma} = p_0^2/
(2 M_{\rm x})$, $B = (e^{E_{\gamma}/k_{\rm B} T}) / (1 - e^{-T_0/T})$,
and $p_0 = (E_{\rm x} \sqrt{\varepsilon_{\rm b}}) / c$ ($E_{\rm x}$ is
the total energy of a ground-state indirect exciton). In contrast with
the evaporative cooling schemes used in atomic optics to remove
high-energy atoms from magnetic traps \cite{Cornell02,Ketterle02}, the
optical evaporation of QW excitons is an inherent process, which deals
with the lowest-energy particles, $0 \leq E \leq E_{\gamma}$, from the
radiative zone. Both signs of $S_{\rm opt}$ can be realized: $S_{\rm opt}
> 0$ ($S_{\rm opt} < 0$), i.e., recombination heating (cooling) of
indirect excitons for $k_{\rm B} T \gtrsim E_{\gamma}$ and $k_{\rm B}
T_0$ ($k_{\rm B} T \ll E_{\gamma}$ and $k_{\rm B} T_0$). In our case
recombination heating occurs: far away from the excitation spot $T$
exceeds $T_{\rm b}$ by about $3$\,mK, i.e., due to relatively high
$T_{\rm b}$, the effect is small. Note that an effective recombination
cooling of particles takes place for $T_{\rm b} \lesssim 0.1$\,K: the
effect is strong and reduces $\tau_{\rm th}$ by $\gtrsim 20\,\%$.

\begin{figure}
\includegraphics*[width=8.8cm]{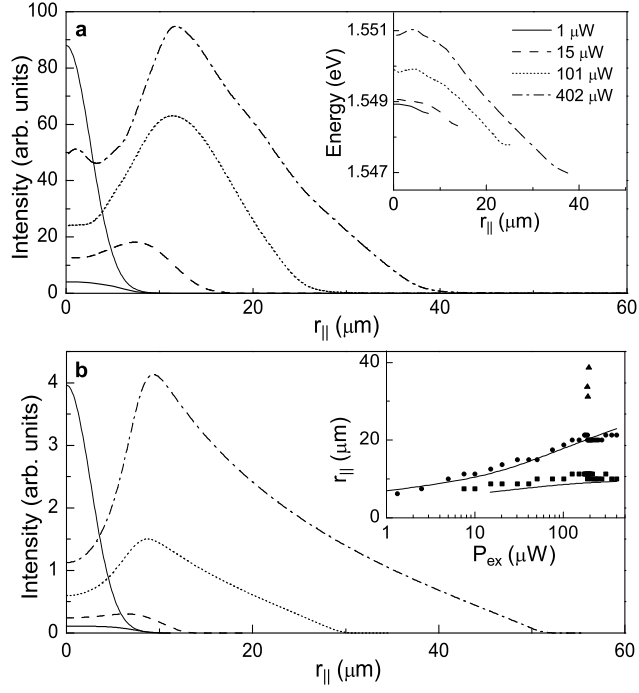}
\caption{ The PL intensity $I_{\rm PL}$, (a) measured and (b)
calculated with Eqs.\,(\ref{diff})-(\ref{opt}), against radius
$r_{\rm \|}$ for four optical excitation powers $P_{\rm ex}$.
The Gaussian profile of the optical excitation, with $\sigma = 3\,\mu$m,
is shown by the thin solid lines. The cryostat temperature
$T_{\rm b} = 1.5$\,K. In numerical evaluations the following parameters
are used: $D_{\rm dp} = 9.6$\,eV, $\tau_{\rm R} = 13$\,ns, $E_{\rm inc}
= 20$\,meV, $M_{\rm x} = 0.215\,m_0$, $d_{\rm QW} = 8$\,nm, and
$d_{\rm z} = 11.5$\,nm. The best fitting parameters are $U_0 = 0.9$\,meV,
$D_{\rm x-imp}^{(2d)} = 60$\,cm$^2$/s, and $C_{\rm x-x} = 15$\,cm$^2$/s.
The upper inset: the energy position of the PL line, $E_{\rm PL} =
E_{\rm PL}(r_{\|})$. The lower inset: the measured (square points) and
calculated (solid line) inner ring radius $r_{\|}^{\rm rg}$ versus
$P_{\rm ex}$. The triangular points refer to the external PL ring. The
measured (circle points) and calculated (solid line) HWHM spatial
extension of the PL signal against $r_{\|}$. }
\label{Fig2}
\end{figure}

Heating of excitons by in-plane drift is given by $S_{\rm d}
= - T_0 u_0 ({\bf v}_{\rm tot}\!\cdot\!\nabla n_{\rm 2d})$
on the r.h.s. of Eq.\,(\ref{therm}), where ${\bf v}_{\rm tot} =
{\bf v}_{\rm diff} + {\bf v}_{\rm drift}$ and ${\bf v}_{\rm diff} = -
(D^{\rm (2d)}_{\rm x}/n_{\rm 2d}) \nabla n_{\rm 2d}$. This heating
mechanism is due to conversion of the mean-field energy into the internal
one. The effect is particularly important for $k_{\rm B} T \gg E_0$
($E_0/k_{\rm B} \simeq 0.4$\,K), when the momentum relaxation time is
much less than $\tau_{\rm th}$.

In our model, the diffusion coefficient $D_{\rm x}^{\rm (2d)} =
(D^{\rm (2d)}_{\rm x-x} D^{\rm (2d)}_{\rm x-imp})/ (D^{\rm
(2d)}_{\rm x-x} + D^{\rm (2d)}_{\rm x-imp})$ has two contributions:
diffusion due to scattering by imperfections (QW impurities and bulk
LA-phonons), with $D^{\rm (2d)}_{\rm x-imp} = D^{\rm (2d)}_{\rm
x-imp}(T_{\rm b})$, and self-diffusion due to exciton-exciton
scattering. The latter channel is important for $n_{\rm 2d} \gtrsim
10^{10}$\,cm$^{-2}$ and $D^{\rm (2d)}_{\rm x-x}$ is approximated by
$D^{\rm (2d)}_{\rm x-x} = C_{\rm x-x}(T/T_0)$ \cite{Ivanov02}. For
$r_{\|}$ far away from the excitation spot, the asymptotic solution
of Eqs.\,(\ref{diff})-(\ref{opt}) yields $I_{\rm PL} \propto
\exp[-(\Gamma_{\rm opt}/ D^{\rm (2d)}_{\rm x-imp})^{1/2}r_{\|}]$. In
contrast, the experimental data show a much more steep decay of the
PL signal and its {\it spatial pinning} at a critical radius
$r^{\rm cr}_{\|} = r^{\rm cr}_{\|}(P_{\rm ex})$ (e.g.,
$r^{\rm cr}_{\|} \simeq 40\,\mu$m for $P_{\rm ex} = 402\,\mu$W,
see Fig.\,2a). We attribute such a behaviour to the
$n_{\rm 2d}$-dependent screening of long-range-correlated
QW disorder by dipole-dipole interacting indirect excitons.
The narrowing effect is illustrated in Fig.\,3a for a particular
realization of the disorder potential $U_{\rm rand}(r_{\|})$.
In order to include the long-range-correlated disorder we use a
{\it thermionic model} \cite{Ivanov02} which operates with
$n_{\rm 2d}$- and $T$-dependent diffusion coefficient
$\tilde{D}^{\rm (2d)}_{\rm x} = D^{\rm (2d)}_{\rm x} \exp[-\,U^{(0)}
/ (k_{\rm B} T + u_0 n_{\rm 2d})]$. Note that for $r_{\|} \simeq
r_{\|}^{\rm cr}$, where the narrowing effect is weak due to small
$n_{\rm 2d}$, a more adequate description of the diffusion and
thermalization kinetics deals with phonon-assisted hopping between
the QW localized states \cite{Baranovskii98,GrassiAlessi00}.

\begin{figure}
\includegraphics*[width=11cm]{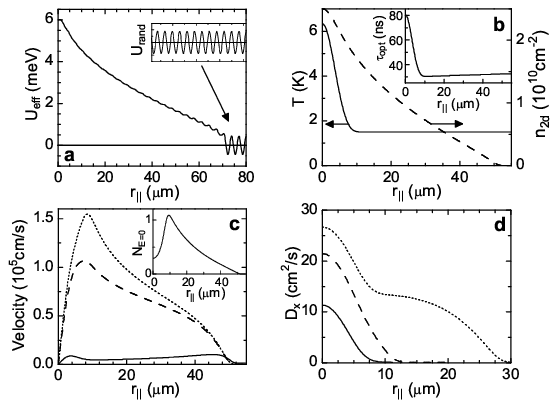}
\caption{ (a) The narrowing effect: screening of the
long-range-correlated disorder potential $U_{\rm rand}({\bf r}_{\|})$
by dipole-dipole interacting indirect excitons. The QW effective
potential $U_{\rm eff}(r_{\|}) = U_{\rm rand}(r_{\|}) + u_0
n_{\rm 2d}(r_{\|})$ calculated with Eqs.\,(\ref{diff})-(\ref{opt})
for harmonic $U_{\rm rand}(r_{\|})$ with $U_0/2 = 0.45$\,meV (see the
inset). (b) $T = T (r_{\|})$ (solid line) and $n_{\rm 2d} =
n_{\rm 2d}(r_{\|})$ (dashed line). Inset: $\tau_{\rm opt} =
\tau_{\rm opt}(r_{\|})$. (c) Diffusion velocity $v_{\rm diff} =
v_{\rm diff}(r_{\|})$ (solid line) and drift velocity $v_{\rm drift}
= v_{\rm drift}(r_{\|})$ without (dashed line) and with (dotted line)
quantum-statistical corrections. Inset: $N_{E=0} = N_{E=0}(r_{\|})$.
Plots (a)-(c) refer to the maximum excitation, $n_{\rm 2d}^{\rm max}
= 2.5 \times 10^{10}$\,cm$^{-2}$ (see the solid lines in Fig.\,2).
(d) Diffusion coefficient $D_{\rm x}^{\rm (2d)} =
D_{\rm x}^{\rm (2d)}(r_{\|})$ for $n_{\rm 2d}^{\rm max} = 0.16 \times
10^{10}$\,cm$^{-2}$ (solid line), $0.44 \times 10^{10}$\,cm$^{-2}$
(dashed line), and $1.27 \times 10^{10}$\,cm$^{-2}$ (dotted line). These
values of $n_{\rm 2d}^{\rm max}$ correspond to the PL signal shown in
Fig.\,2. }
\label{Fig3}
\end{figure}

In order to model the experimental results within the developed
microscopic picture, we solve Eqs.\,(\ref{diff})-(\ref{opt}) numerically
for a stationary, cylindrically-symmetric optical excitation profile,
so that the generation rate $\Lambda({\bf r}_{\|},t) \equiv
\Lambda(r_{\|}) \propto P_{\rm ex} \exp(-r_{\|}^2/\sigma^2)$.
The best fit for the experimental data plotted in Fig.\,2a yields $U_0
= 0.9$\,meV, $D_{\rm x-imp}^{\rm (2d)} = 60$\,cm$^2$/s, and $C_{\rm x-x}
= 15$\,cm$^2$/s. The calculated spatial profile of the PL signal,
$I_{\rm PL} = I_{\rm PL}(r_{\|})$, is shown in Fig.\,2b for various
pump powers $P_{\rm ex}$. While the density profile $n_{\rm 2d}
= n_{\rm 2d}(r_{\|})$ always has a bell-like shape (see Fig.\,3b and
the inset of Fig.\,2a), with increasing $P_{\rm ex}$ the inner
PL ring develops in the $I_{\rm PL}$-profile. This is in a complete
agreement with the observations. The inner ring has a nearly classical
origin, and arises due to heating of indirect excitons by the
optical excitation [$S_{\rm pump}$ term in Eq.\,(\ref{therm})]:
with increasing $r_{\|}$ the exciton temperature $T$ rapidly decreases
towards $T_{\rm b}$ (see Fig.\,3b); as a result, the optical lifetime
$\tau_{\rm opt} = 1/\Gamma_{\rm opt}$ decreases too (see the inset of
Fig.\,3b), giving rise to a local increase of $I_{\rm PL}(r_{\|})$ at
$r_{\|} = r^{\rm rg}_{\|}$. Thus the inner ring is a spatial
counterpart of the PL-jump observed in the time-resolved experiments
\cite{Butov01}. Our numerical simulations also reproduce the observed
increase of $r_{\|}^{\rm rg}$ and the spatial extension of the PL area
(HWHM of the signal) with increasing $P_{\rm ex}$ (see the inset
of Fig.\,2b).

The finding of the fitting parameters, which refer to the total
diffusion coefficient $\tilde{D}^{\rm (2d)}_{\rm x}$, is {\it complex},
i.e., we fit all the curves plotted in Fig.\,2a [$I_{\rm PL} =
I_{\rm PL}(r_{\|})$ and $E_{\rm PL} = E_{\rm PL}(r_{\|})$ for various
$P_{\rm ex}$] by using the same values of $U_0$,
$D_{\rm x-imp}^{\rm (2d)}$, and $C_{\rm x-x}$. The blue shift
$\delta_{\rm PL} \geq 0$ of the PL energy $E_{\rm PL}$ is due to the
mean-field interaction energy of indirect excitons, $\delta_{\rm PL} =
E_{\rm PL} - E_{\rm x} = u_0 n_{\rm 2d}$ (see Figs.\,1e-1f and the inset
of Fig.\,2a). Thus we use the measured $\delta_{\rm PL}$ to estimate
$n_{\rm 2d}^{\rm max}(r_{\|} = 0)$, and therefore the generation
rate $\Lambda(r_{\|} = 0)$, necessary for numerical modelling
with Eqs.\,(\ref{diff})-(\ref{opt}). The amplitude of the disorder
potential $U_0$ determines the steepness of $I_{\rm PL}(r_{\|} >
r_{\|}^{\rm rg})$ and dependence $r_{\|}^{\rm cr} =
r_{\|}^{\rm cr}(P_{\rm ex})$ of the PL pinning radius. In turn,
$D_{\rm x-imp}^{(2d)}$ and $C_{\rm x-x}$ determine the ring contrast
and $r_{\|}^{\rm rg} = r_{\|}^{\rm rg}(P_{\rm ex})$ dependence. The
total diffusion coefficient $\tilde{D}^{\rm (2d)}_{\rm x} =
\tilde{D}^{\rm (2d)}_{\rm x}(r_{\|})$ is plotted in Fig.\,3d.

In Fig.\,3c we show that for the small size excitation spot ($\sigma
\simeq 3\,\mu$m) used in the experiments, the drift velocity
$v_{\rm drift}$, due to the gradient of the mean-field interaction
energy $u_0n_{\rm 2d}(r_{\|})$, is much larger than the diffusion
velocity $v_{\rm diff}$. The total velocity has a maximum value
$v^{\rm max}_{\rm tot}(r_{\|} \simeq r_{\|}^{\rm rg}) \simeq 1.5 \times
10^5$\,cm/s for $n_{\rm 2d}^{\rm max}(r_{\|}=0) \simeq 2.5 \times
10^{10}$\,cm$^{-2}$ (see Fig.\,3c). Note that in our case the
mean-field energy gradient $u_0 |\nabla n_{\rm 2d}(r_{\|} \simeq
r_{\|}^{\rm rg} )| \simeq 1.6$\,eV/cm exceeds the maximum strain-induced
gradient $|\nabla U| \simeq 0.4$\,eV/cm used in the experiments
\cite{Tamor80,Trauernicht84}.

Finally, we emphasize that in our experiments, which deal with the
cryostat temperature $T_{\rm b} = 1.5$\,K, nonclassical occupation
numbers of modest values, $N^{\rm max}_{E=0}(r_{\|} \simeq
r_{\|}^{\rm rg}) \simeq 1$, build up at the position of the inner
ring (see the inset of Fig.\,3c). As illustrated in Fig.\,3c, in
this case the quantum statistical corrections, e.g., to $v_{\rm diff}$
and to the Einstein relationship, are about 35\% and therefore cannot
be neglected. Nonclassical statistics occurs at the position of the
inner ring, where the exciton gas is already cold but still dense.
Furthermore, for $T_{\rm b} \sim 0.1$\,K  (not yet realized in an
optical imaging experiment) numerical modelling with
Eqs.\,(\ref{diff})-(\ref{opt}) gives well-developed Bose-Einstein
statistics with $N^{\rm max}_{E=0}(r_{\|} \simeq r_{\|}^{\rm rg}) \gg 1$.

We appreciate valuable discussions with L.\,V. Keldysh, L.\,S.
Levitov, L. Mouchliadis, and B.\,D. Simons. Support of this work by
EU RTN Project HPRN-2002-00298 is gratefully acknowledged.

\end{document}